\documentclass[sigconf,authorversion]{acmart}
\usepackage{makecell}
\usepackage{ragged2e}
\usepackage{longtable}
\usepackage[normalem]{ulem}
\usepackage{graphicx}
\usepackage{booktabs}
\usepackage{threeparttablex}
\usepackage[dvipsnames]{xcolor}
\usepackage{amsmath}
\usepackage{pifont}

\usepackage[width=.99\textwidth]{caption}
\usepackage{rotating}
\usepackage{pdflscape}
\usepackage{float}
\usepackage{multirow}
\usepackage{amsmath}
\usepackage{rotating}
\usepackage{tikz}
\usepackage{dblfloatfix}
\usepackage{caption}
\usepackage{subcaption}
\usepackage[ruled]{algorithm2e}
\usepackage{listings}
\usepackage{xcolor,pifont}
\usepackage{booktabs}
\usepackage{multirow}
\usepackage{siunitx}

\AtBeginDocument{%
  }

\setcopyright{acmlicensed}
\copyrightyear{2026}
\acmYear{2026}
\acmDOI{XXXXXXX.XXXXXXX}
\acmConference[ICFNDS '26]{}{Nov 04--06,2026}{Izmir, Turkey}
\acmISBN{978-1-4503-XXXX-X/2018/06}

\begin{document}

\title{OVS Meets PQ-TLS: Exploring Post-Quantum TLS for SDN's Southbound API}

\author{Majd Latah}
\affiliation{%
  \institution{Ozyegin University}
  \city{Istanbul}
  \country{Turkey}}
\email{majd.latah@ozu.edu.tr}

\author{Kubra Kalkan}
\affiliation{%
  \institution{Ozyegin University}
  \city{Istanbul}
  \country{Turkey}}
\email{kubra.kalkan@ozyegin.edu.tr}

\renewcommand{\shortauthors}{Majd Latah and Kubra Kalkan}

\begin{abstract}
 Software-defined networking (SDN) is a novel networking paradigm that enables network programmability and centralized control for network devices. The southbound application programming interface (API) is used to control and manage the underlying data plane devices. The existing southbound API relies on TLS with legacy cryptographic algorithms such as RSA and ECDSA. In this paper, we explore the performance of the southbound API with post-quantum TLS (PQ-TLS) support. We present a proof-of-concept of using PQ-TLS in SDN's southbound API. We study the performance of pure and hybrid PQ-TLS modes across different security levels and compare them with legacy TLS in terms of latency and CPU utilization. We also compare the performance of different post-quantum signature and key establishment schemes.
\end{abstract}


\maketitle

\section{Introduction}
SDN consists of three main planes: 1) data plane 2) control plane, and 3) application plane \cite{survey2Wenfeng,survey3}. SDN planes are connected through two main Application Programming Interfaces (APIs) \cite{survey2Wenfeng,survey3}:
\begin{itemize}
  \item \textbf{Southbound API:} an interface between the SDN controller and a corresponding SDN switch.
  \item \textbf{Northbound API:} an interface between the SDN applications and a corresponding SDN controller.
\end{itemize}

\begin{figure}[t!]
\centering
\includegraphics[scale=0.45]{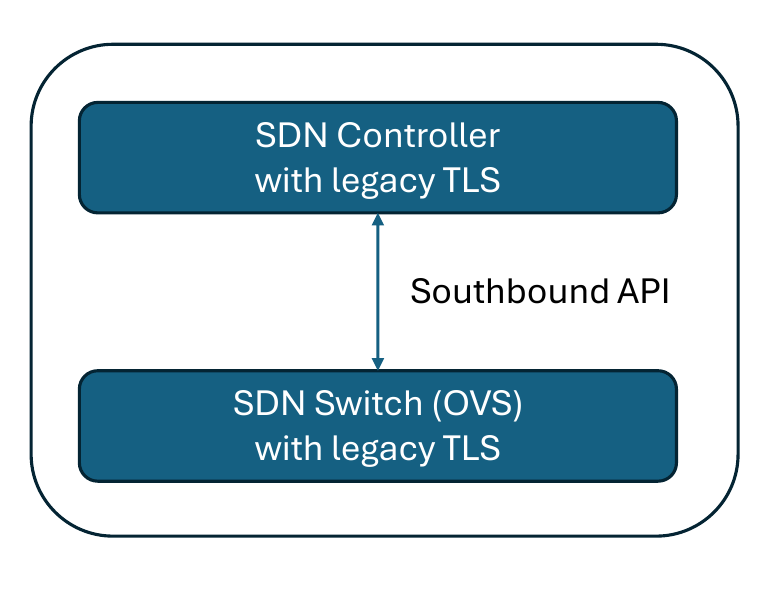}
\caption{SDN's southbound API protected using legacy TLS}
\label{Southbound-API}
\end{figure}

Our work focuses on the security of the southbound API. In this context, SDN uses the Transport Layer Security (TLS) protocol to secure its southbound API (See Fig. \ref{Southbound-API}). Existing approaches based on legacy TLS do not provide security against post-quantum (PQ) adversaries. It is worth noting that the cryptographic algorithms used in legacy TLS, such as ECDSA and RSA, do not provide security against quantum attacks. In this regard, NIST defined five security categories \cite{NIST-Sec-Levels} for post-quantum alternatives. PQ-TLS relies on cryptographic algorithms that claim to satisfy those security levels. In this work, we explore PQ-TLS for protecting the southbound API in SDN. We focus on a well-known SDN switch (i.e., OVS), which is a software switch widely used in SDN. We present a proof of concept in which PQ-TLS is compared with legacy TLS in terms of handshake latency and CPU utilization. In addition, we compare the performance of different PQ signature and key establishment schemes for securing the southbound API in SDN environments.

\section{Related work}
The southbound API of SDN relies on the TLS protocol to protect the communication channel between the controller and SDN switches. Several approaches focus on a secure southbound API, such as \cite{durner2015cost,yigit2019secured,liyanage2014securing,latah2020dpsec}. In \cite{durner2015cost}, the authors study packet-in delays of SDN switches when TLS encryption is used. In \cite{yigit2019secured}, the authors focus on improving the security of SDN by combining Access Control Lists (ACLs) with TLS hardening. In \cite{liyanage2014securing}, the authors propose a solution based on Host Identity Protocol (HIP) to secure the communication channel in software-defined mobile networks. In \cite{latah2020dpsec}, the authors introduce a blockchain-based solution to authenticate SDN switches and SDN hosts. 
In \cite{rivera2025leveraging}, the authors suggest a combination of blockchain and post-quantum cryptography for securing cross-domain SDN environments. Adding blockchain results in additional complexity and is considered more suitable solution for cross-domain SDNs \cite{latah2020dpsec,rivera2025leveraging}.

In \cite{buruaga2025hybrid}, the authors propose a solution that integrates Quantum Key Distribution (QKD) and post-quantum cryptography (PQC) in SDN integration. Our work provides a more practical solution than \cite{buruaga2025hybrid} since it does not relay on QKD. Furthermore, our solution is integrated directly into existing SDN components such as OVS switches. 

In \cite{mrinal2025performance}, the authors explored the integration of post-quantum algorithms for securing SDN environments. However, the paper did not provide a comprehensive and systematic comparison across legacy, hybrid, and pure PQ-TLS. Our work covers three main security levels across legacy, hybrid and pure PQ-TLS. We also evaluate other alternatives for PQ algorithms. Furthermore, the paper did not include detailed information on switch integration, which is a key focus of our work. Other works, such as \cite{sosnowski2023performance}, focus on providing PQ-TLS measurements for non-SDN environments. Our work, on the other hand, focuses on SDN environments with emphasis on the southbound API.

\section{Southbound API with PQ-TLS}
\begin{figure}[h!]
\centering
\includegraphics[scale=0.4]{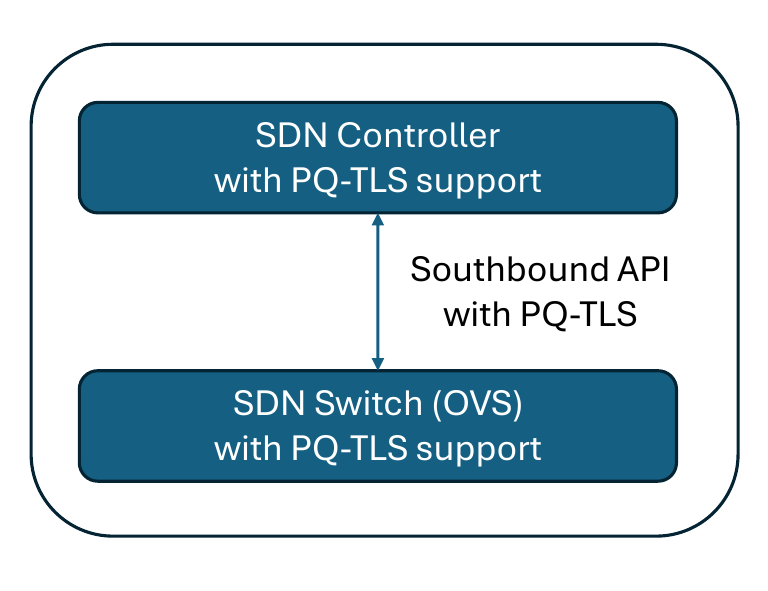}
\caption{SDN's southbound API protected using PQ-TLS}
\label{Southbound-API-PQ}
\end{figure}

To achieve PQ-resiliency for the southbound API, SDN controllers and SDN switches need to provide support for PQ-TLS. Therefore, both SDN components need to use PQ-TLS to communicate with each other (see Fig. \ref{Southbound-API-PQ}).
In this paper, we consider the following modes:

\begin{itemize}
  \item \textbf{Pure PQ-TLS mode:} uses only post-quantum cryptographic algorithms. This mode uses ML-DSA for digital signatures and ML-KEM for key establishment. Both ML-DSA and ML-KEM are module-lattice-based algorithms standardized by NIST. 
  \item \textbf{Hybrid PQ-TLS mode:} uses a combination of classical and post-quantum algorithms. This mode combines ECDSA with ML-DSA for digital signatures and ECDH with ML-KEM for key establishment.
\end{itemize}

\subsection{SDN switches with PQ-TLS support}
The SDN data plane, which consists mainly of OVS switches\footnote{OVS, \url{https://github.com/openvswitch/ovs}}, relies on PQ-TLS to communicate with the SDN controller. To achieve PQ-TLS support for the pure PQ-TLS mode, the switch uses the new OpenSSL library with PQ-TLS support. In addition, to achieve PQ-TLS support for hybrid PQ-TLS mode, we use a combination of OpenSSL \footnote{OpenSSL,  \url{https://github.com/openssl/openssl}} and liboqs \footnote{Liboqs, \url{https://github.com/open-quantum-safe/liboqs}} from the Open Quantum Safe (OQS) project. Note that oqs-provider \footnote{Oqs-provider, https://github.com/open-quantum-safe/oqs-provider} is used to enable liboqs in OpenSSL. The source code of the switch is compiled with the new OpenSSL library with PQ-TLS support.

\subsection{SDN controller with PQ-TLS support}
The SDN controller uses PQ-TLS to create a secure channel to communicate with the data plane. For this part, we integrate PQ-TLS into an existing SDN controller (Ryu). We choose Ryu due to its simplicity, which allows rapid testing of new functionalities in the SDN environment. Both pure and hybrid modes are also supported at the controller side through OpenSSL, liboqs and oqs-provider. The controller also uses a Python C extension module to call the PQ-enabled OpenSSL.
In addition, the controller runs a L2 learning switch application.

\begin{figure}[b!]
\includegraphics[scale=0.58]{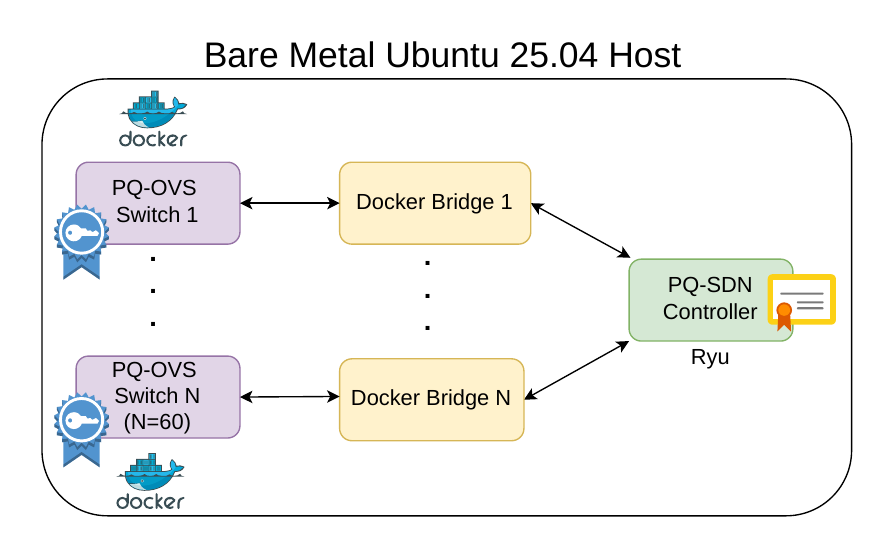}
\caption{Our testbed}
\label{Testbed}
\end{figure}

\begin{figure*}[b!]
\centering
\begin{subfigure}[b]{0.33\textwidth}
    \centering
    \includegraphics[width=\textwidth]{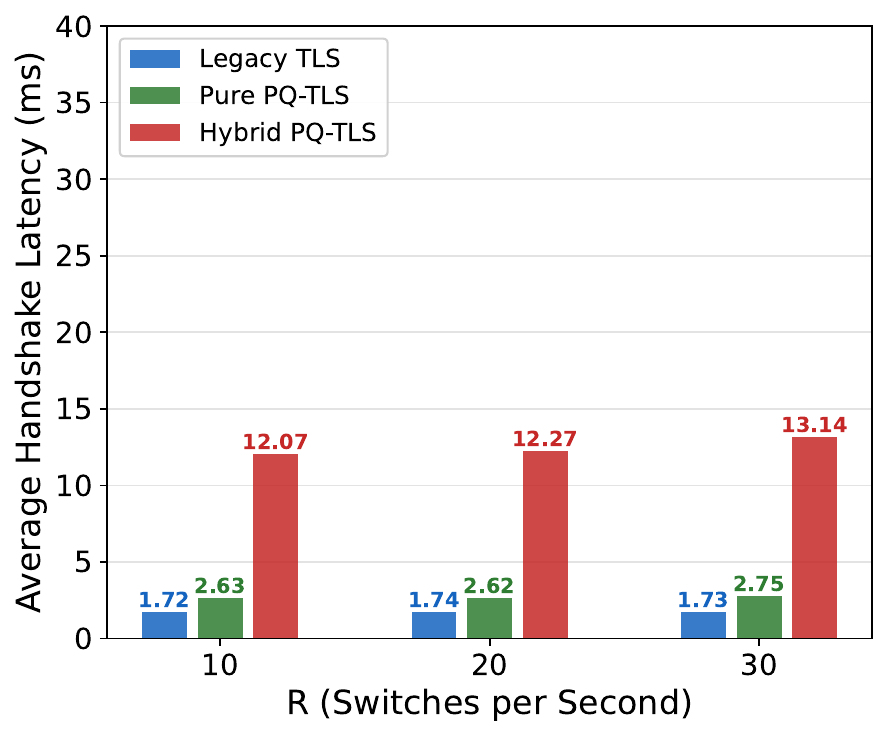}
    \caption{\footnotesize L1}
    \label{Latency:Sub1}
\end{subfigure}
\hfill
\begin{subfigure}[b]{0.33\textwidth}
    \centering
    \includegraphics[width=\textwidth]{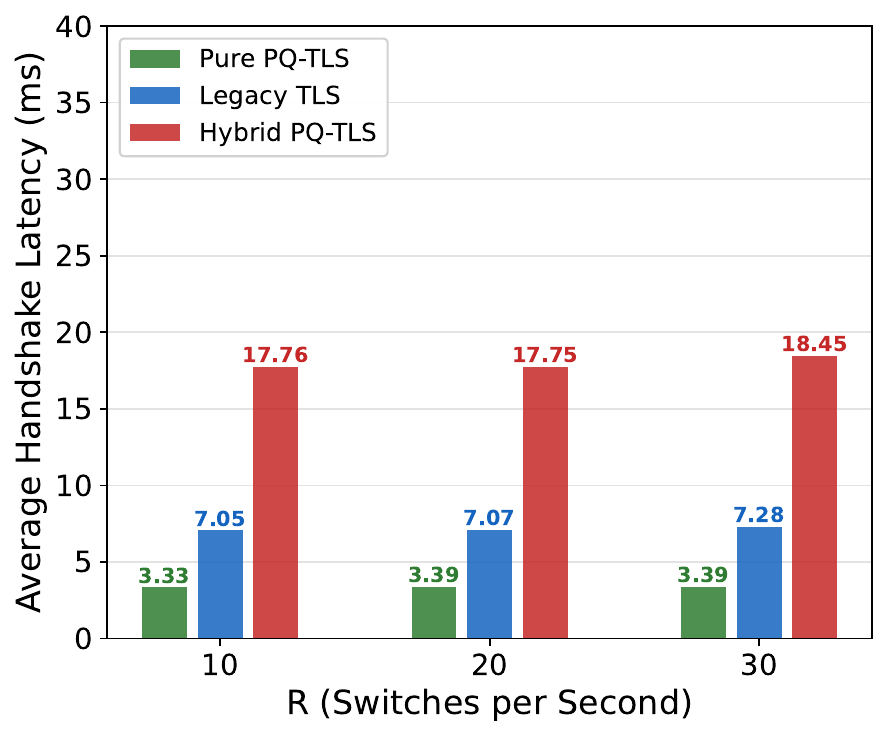}
    \caption{\footnotesize L3}
    \label{Latency:Sub2}
\end{subfigure}
\hfill
\begin{subfigure}[b]{0.33\textwidth}
    \centering
    \includegraphics[width=\textwidth]{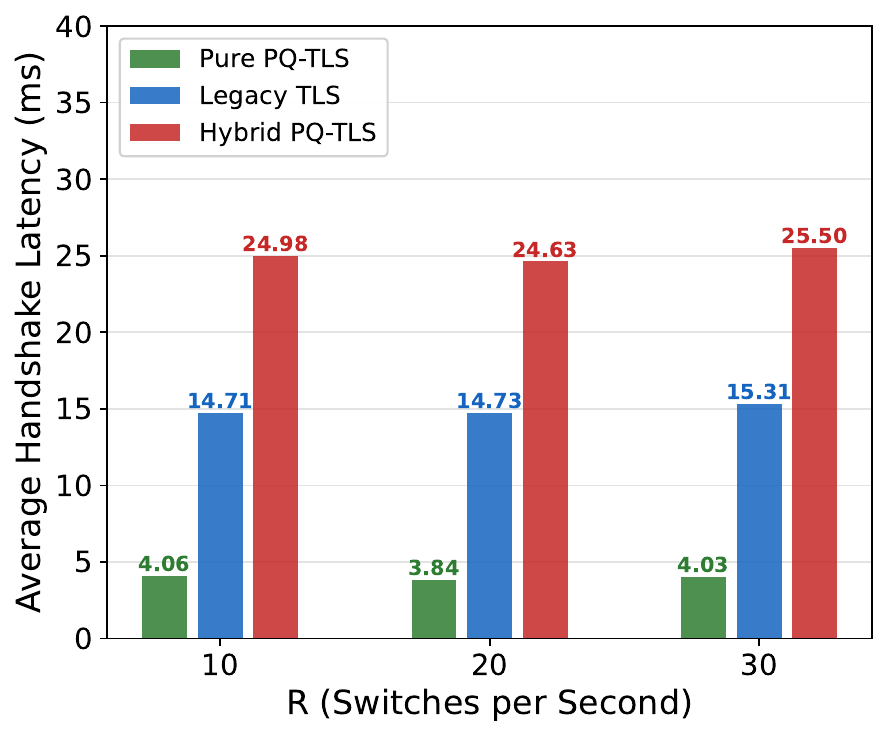}
    \caption{\footnotesize L5}
    \label{Latency:Sub3}
\end{subfigure}
\captionsetup{justification=centering}
\caption{Handshake latency for legacy TLS and PQ-TLS in the data plane across different security levels}
\label{Latency:Single}
\end{figure*}

\begin{figure*}[b!]
\centering
\begin{subfigure}[b]{0.33\textwidth}
    \centering
    \includegraphics[width=\textwidth]{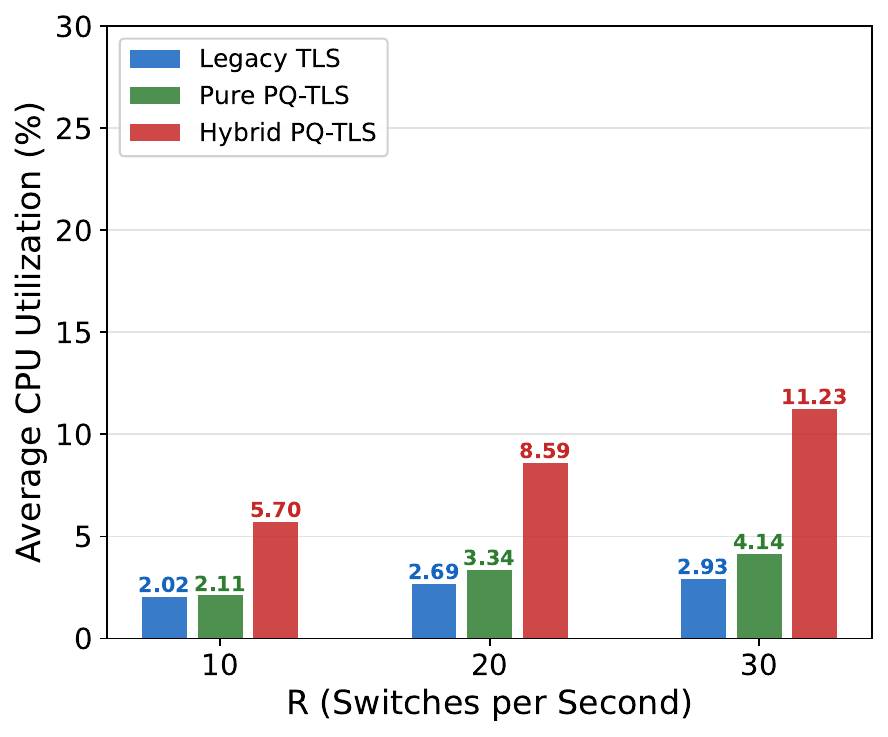}
    \caption{\footnotesize L1}
    \label{CPU:Sub1}
\end{subfigure}
\hfill
\begin{subfigure}[b]{0.33\textwidth}
    \centering
    \includegraphics[width=\textwidth]{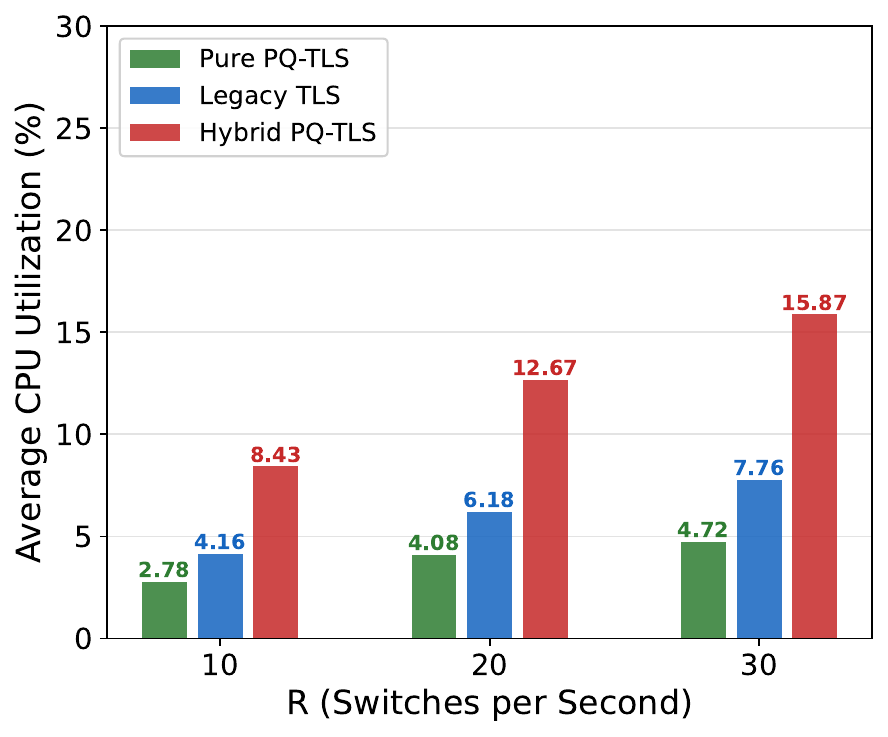}
    \caption{\footnotesize L3}
    \label{CPU:Sub2}
\end{subfigure}
\hfill
\begin{subfigure}[b]{0.33\textwidth}
    \centering
    \includegraphics[width=\textwidth]{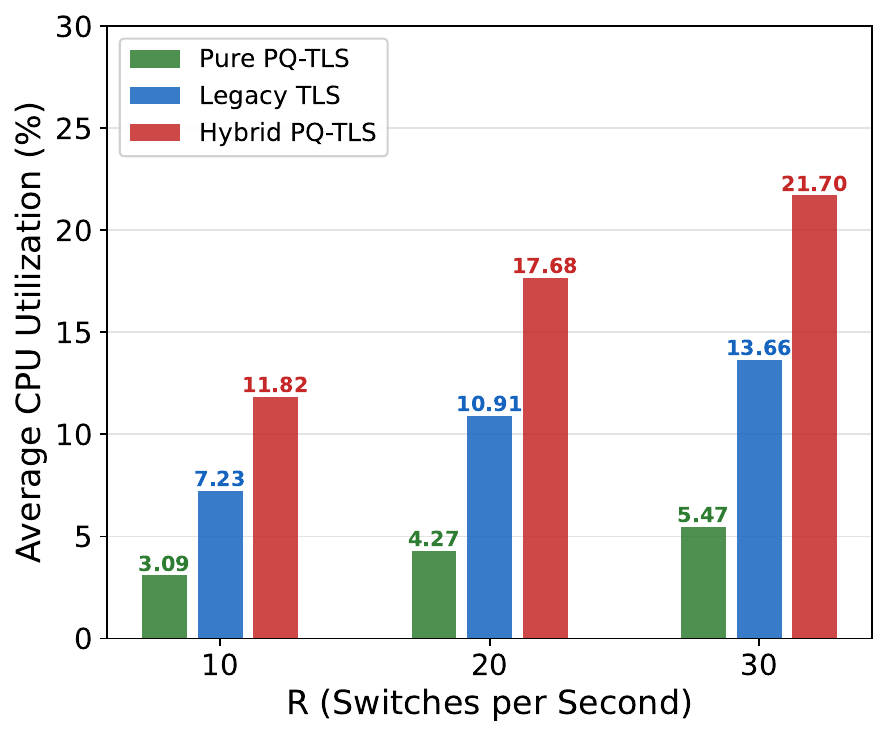}
    \caption{\footnotesize L5}
    \label{CPU:Sub3}
\end{subfigure}
\caption{CPU utilization for legacy TLS and PQ-TLS in the control plane across different security levels}
\label{CPU-Usage}
\end{figure*}

\section{Experimental Work}
The experimental work is conducted on a bare-metal machine with the Ubuntu 25.04 operating system. The machine has an AMD Ryzen 7 (7840HS) CPU and 16 GB of RAM. Our testbed is shown in Fig. \ref{Testbed}. The details of our experimental setup are shown in Table \ref{Setup-table}. During our AI-assisted analysis using Claude Code, we noticed that the TLS socket is not monitored by the switch event loop, causing additional latency during the handshake process. This issue is fixed by registering the SSL socket in the corresponding event loop.

\begin{table}[h!]
\centering
\caption{Experimental setup}
\label{tab:testbed}
\begin{tabular}{|l|l|}
\hline
\textbf{Setting} & \textbf{Description} \\ \hline
Operating system   & Ubuntu 25.04 \\    \hline
SDN controller     & Ryu \\   \hline
OpenSSL        & 4.0.0-dev \\ \hline
Liboqs         & 0.15.0 \\ \hline
SDN switch   & PQ Open vSwitch (OVS) 3.6.90 \\      \hline
Containerization   & One container per switch \\
                   \hline
Network isolation  & Dedicated L2 network per switch \\    \hline
Number of switches  & $N = 60$  \\                       \hline
Switch arrival rates & $R \in \{10, 20, 30\}$ switches per second \\         \hline
\end{tabular}
\label{Setup-table}
\end{table}

\begin{table}[b!]
\centering
\caption{Parameters used for TLS configuration}
\label{tab:tls-parameters}
\begin{tabular}{|l|l|l|l|}
\hline
\textbf{Level} & \textbf{TLS Mode} & \textbf{Digital Sig} & \textbf{Key Establishment} \\
\hline
\multirow{3}{*}{L1$^{*}$} & Legacy  &  p256          & secp256r1       \\
                           & Pure PQ    & MLDSA44       & MLKEM512       \\
                           & Hybrid PQ & p256\_mldsa44 & p256\_mlkem512  \\
\hline
\multirow{3}{*}{L3}       & Legacy  & p384        & secp384r1       \\
                           & Pure PQ   & MLDSA65       & MLKEM768        \\
                           & Hybrid PQ & p384\_mldsa65 & p384\_mlkem768 \\
\hline
\multirow{3}{*}{L5}       & Legacy & p521          & secp521r1       \\
                           & Pure PQ  & MLDSA87       & MLKEM1024       \\
                           & Hybrid PQ  & p521\_mldsa87 & p521\_mlkem1024 \\
\hline
\end{tabular}

\vspace{3pt}
\raggedright\footnotesize $^{*}$For digital signature, since no NIST Level 1 variant for ML-DSA, we choose the lowest variant MLDSA44, which is NIST Level 2.
\end{table}

In our setup, each switch starts an independent TLS 1.3 session with the controller. Therefore, each OVS switch runs in a separate Docker container and has its own certificate and cryptographic keys. Each container is associated with a dedicated Docker bridge, which provides connectivity to the SDN controller. 

The network consists of $N = 60$ switches. The switch arrival rate ($R$) is varied over three values: 10, 20, and 30 switches per second. The SDN controller runs directly on the host machine. We use CPU affinity to bind the controller to a single CPU core. 

We consider both pure and hybrid PQ-TLS modes. The parameters for TLS configuration are shown below in Table \ref{tab:tls-parameters}. We consider three NIST security levels: L1, L3, and L5 (128, 192, and 256 bit). The symmetric cipher suite is \textit{TLS\_AES\_256\_GCM\_SHA384}. It is worth noting that each experiment is repeated 10 times.

\subsection{Comparing handshake latency and CPU utilization for legacy TLS and PQ-TLS:}
The handshake latency is reported in Fig. \ref{Latency:Single}. The results show that using a higher security level results in higher latency. For L1, legacy TLS shows the least latency. However, for L3 and L5, pure PQ-TLS demonstrates the lowest latency. The results confirm that classical algorithms constrain the performance of hybrid algorithms at higher security levels, which is consistent with \cite{sosnowski2023performance}. The hybrid PQ-TLS mode shows the highest latency across different security levels, which is expected since it combines both classical and quantum-resistant algorithms. We also measure CPU utilization for both legacy TLS and PQ-TLS at the controller side. The results for CPU utilization are shown in Fig. \ref{CPU-Usage}. We observe that CPU utilization increases linearly with switch arrival rate ($R$). For L1, legacy TLS shows the least CPU utilization, whereas for high security levels (L3 and L5) pure PQ-TLS shows the lowest utilization.

\begin{figure*}[b!]
\centering
\begin{subfigure}[b]{0.48\textwidth}
    \centering
    \includegraphics[width=\textwidth]{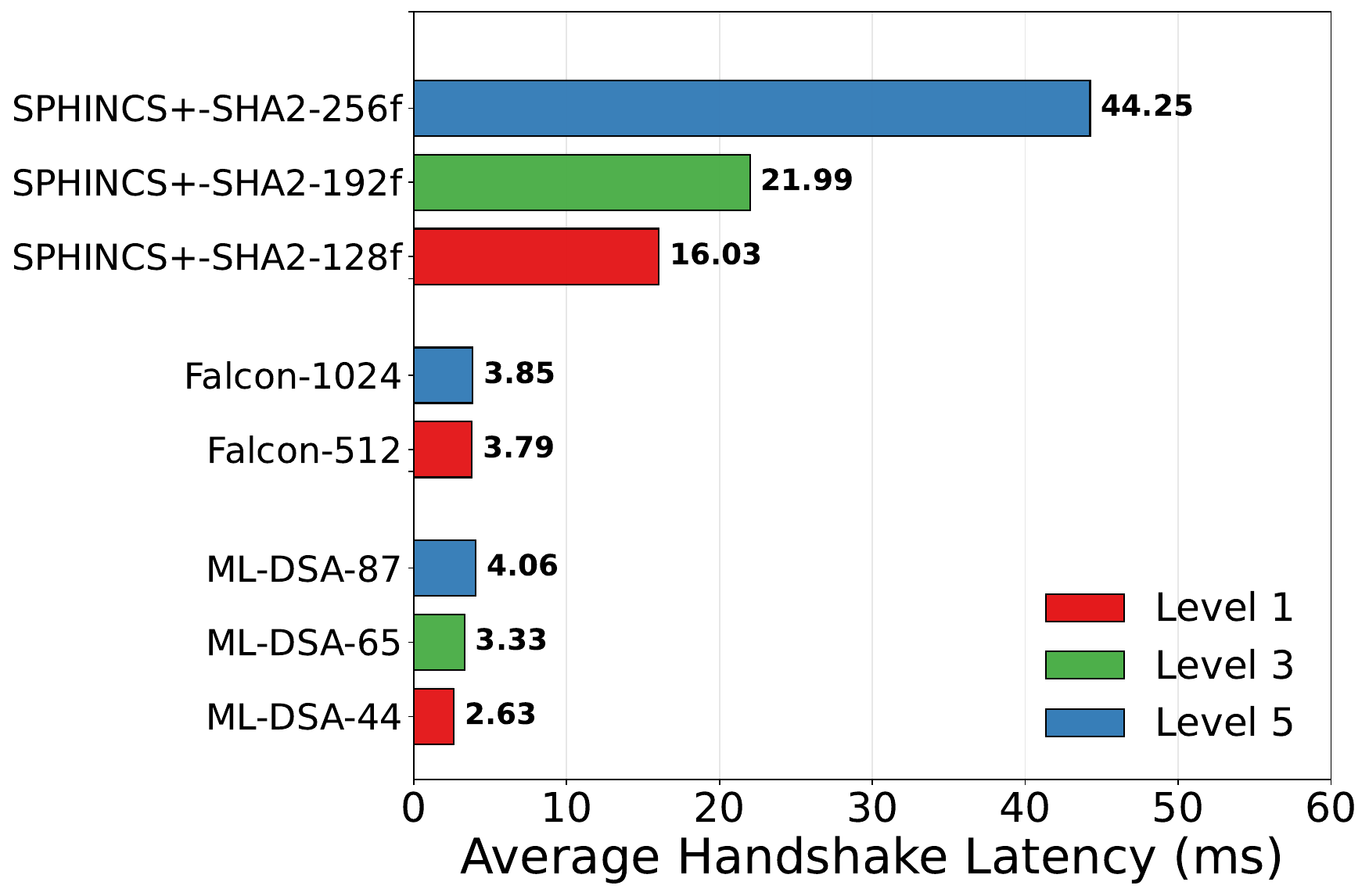}
    \caption{\footnotesize Handshake latency in data plane}
    \label{CPU:Sub1}
\end{subfigure}
\hfill
\begin{subfigure}[b]{0.48\textwidth}
    \centering
    \includegraphics[width=\textwidth]{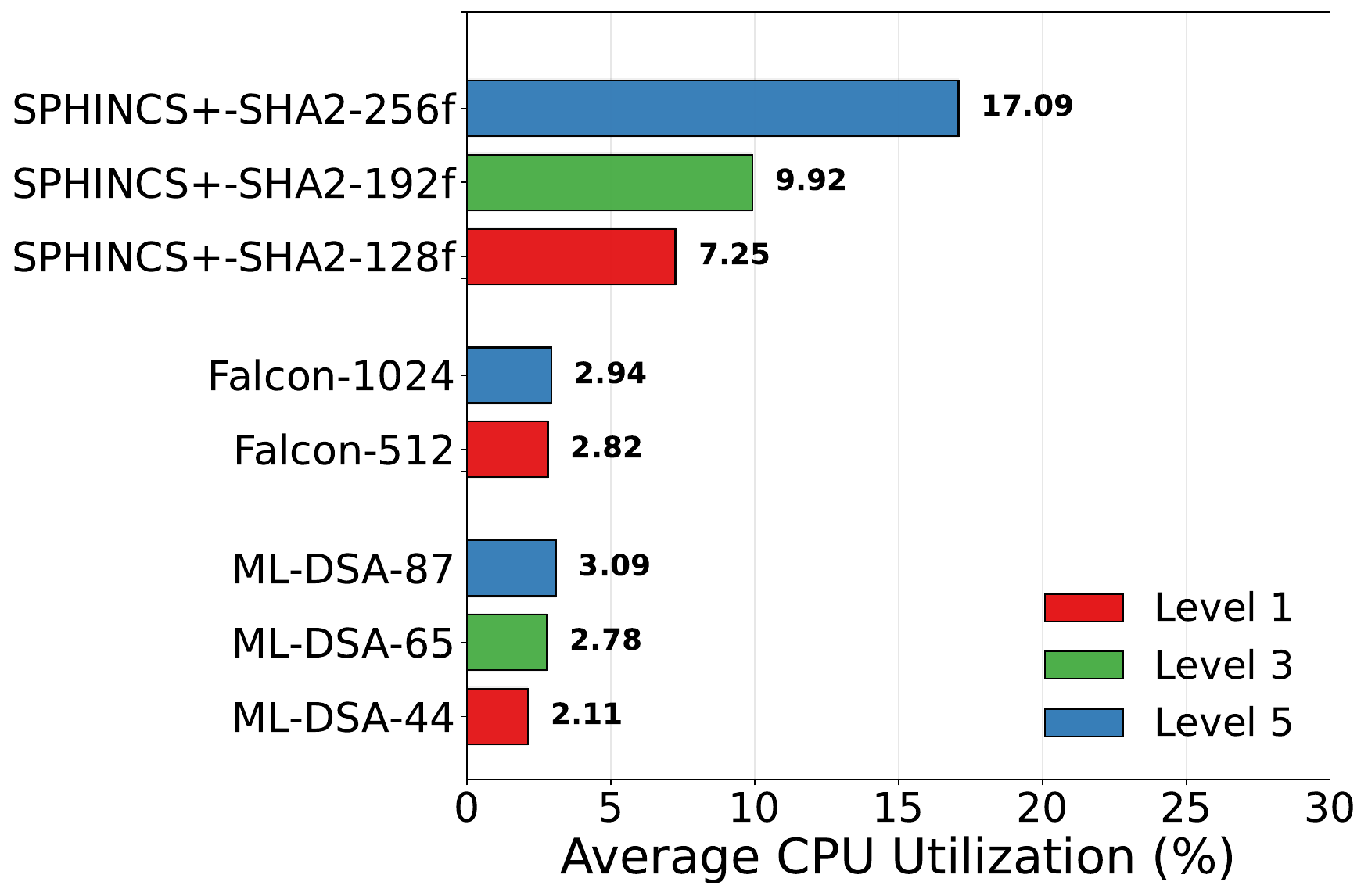}
    \caption{\footnotesize CPU utilization in control plane}
    \label{CPU:Sub2}
\end{subfigure}
\caption{Comparing ML-DSA with SPHINCS+ and Falcon 
(ML-KEM is used for key establishment)}
\label{SPHINCS}
\end{figure*}

\begin{figure*}[b!]
\centering
\begin{subfigure}[b]{0.49\textwidth}
    \centering
    \includegraphics[width=\textwidth]{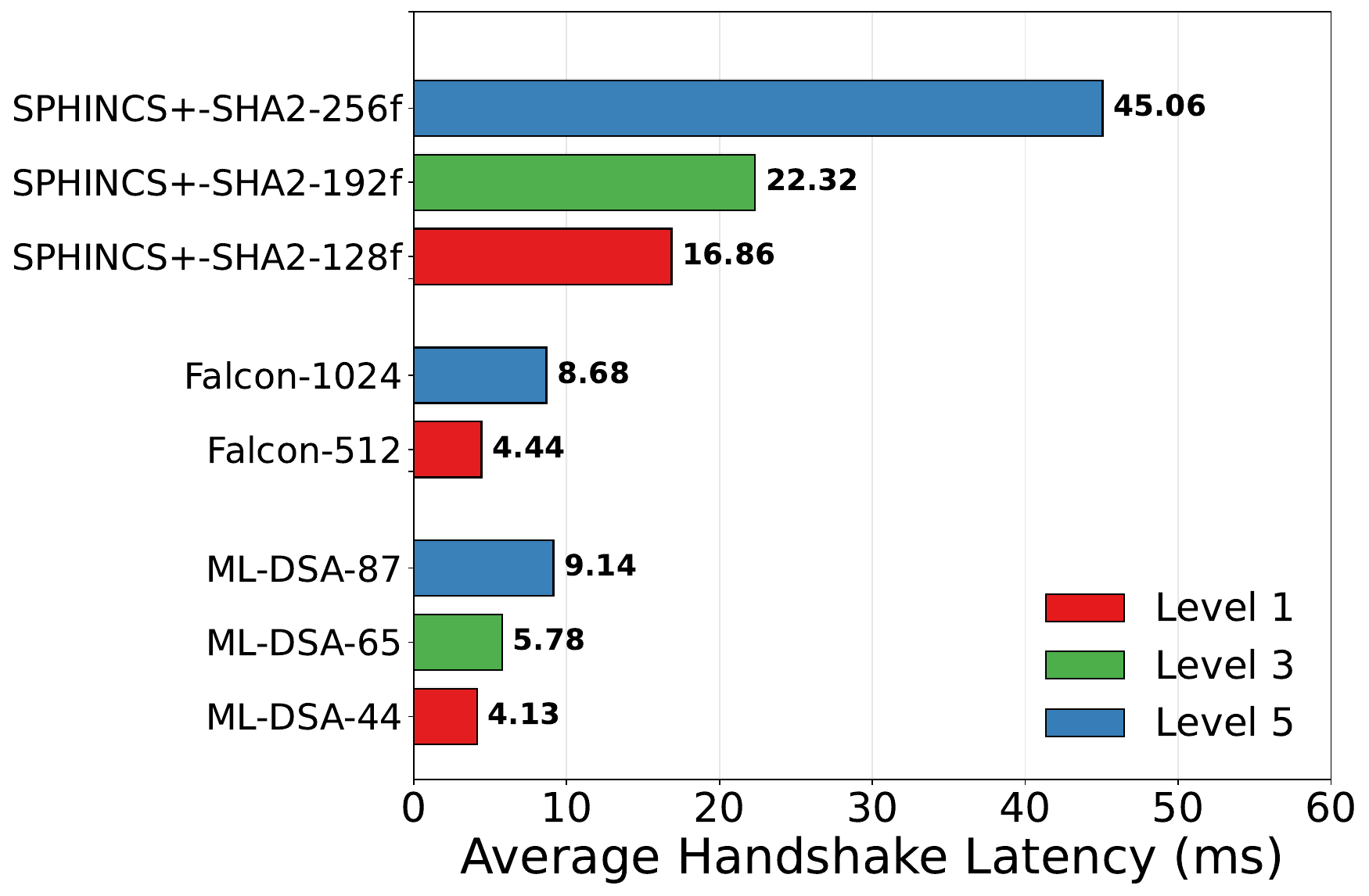}
    \caption{\footnotesize Handshake latency in data plane}
    \label{CPU:Sub1}
\end{subfigure}
\hfill
\begin{subfigure}[b]{0.49\textwidth}
    \centering
    \includegraphics[width=\textwidth]{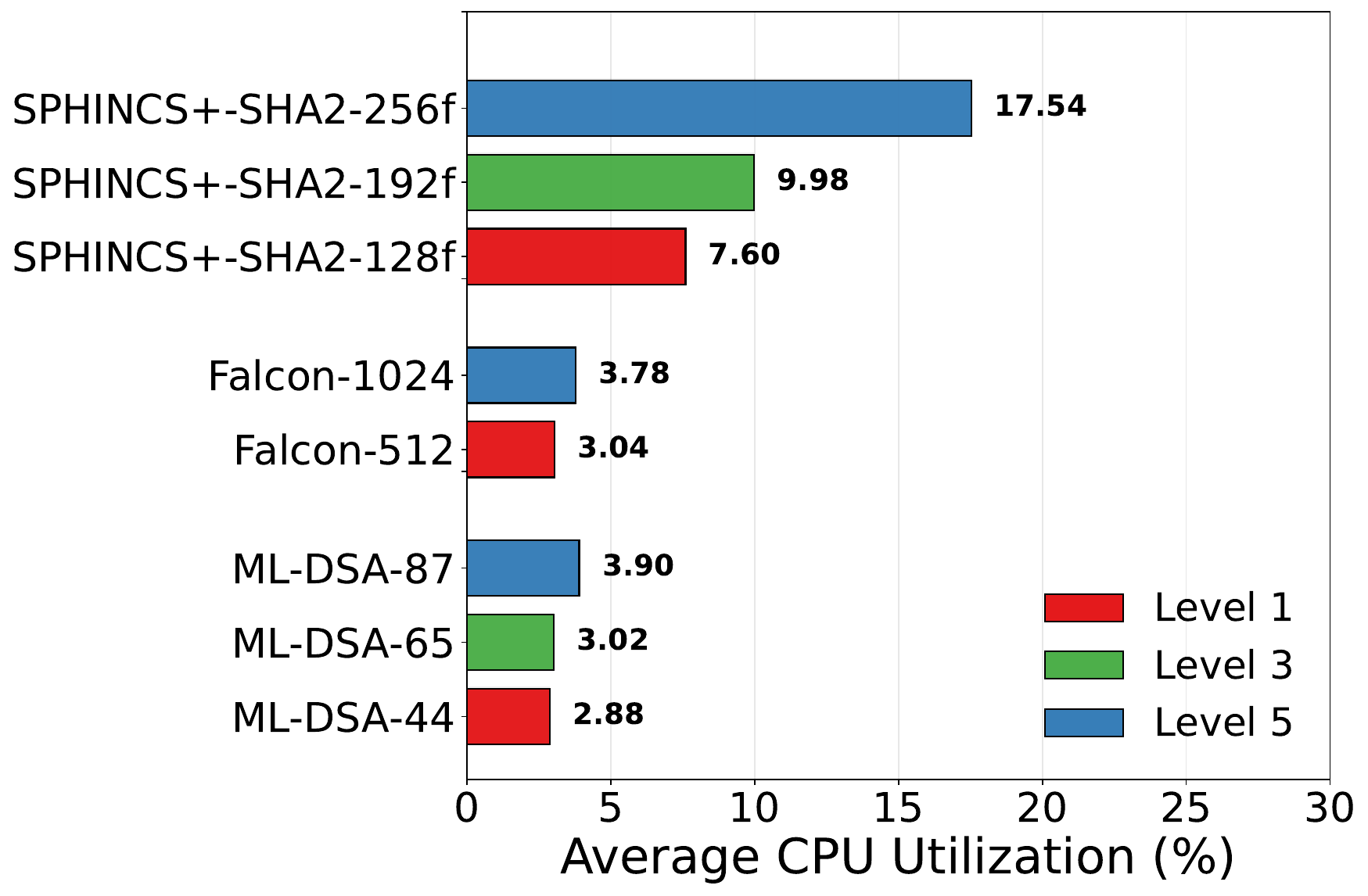}
    \caption{\footnotesize CPU utilization in control plane}
    \label{CPU:Sub2}
\end{subfigure}
\caption{Comparing ML-DSA with SPHINCS+ and Falcon (BIKE is used for key establishment)}
\label{BikeFig}
\end{figure*}

\subsection{Comparing with other PQ digital signatures and key establishment schemes:}
For digital signatures, we consider both SPHINCS+ and Falcon. SPHINCS+ is a hash-based post-quantum digital signature scheme, which is also standardized by NIST as SLH-DSA. Falcon is a lattice-based post-quantum signature scheme, which is also standardized by NIST as FN-DSA. For key establishment, we consider BIKE, FrodoKEM and HQC. Note that Falcon has only two security levels (L1 and L5). We use the fast variants of SPHINCS+. We fix the switch arrival rate at $R=10$. We compare the performance of ML-DSA with SPHINCS+ and Falcon across different security levels and different key establishment schemes.

\begin{figure*}[t!]
\centering
\begin{subfigure}[b]{0.445\textwidth}
    \centering
    \includegraphics[width=\textwidth]{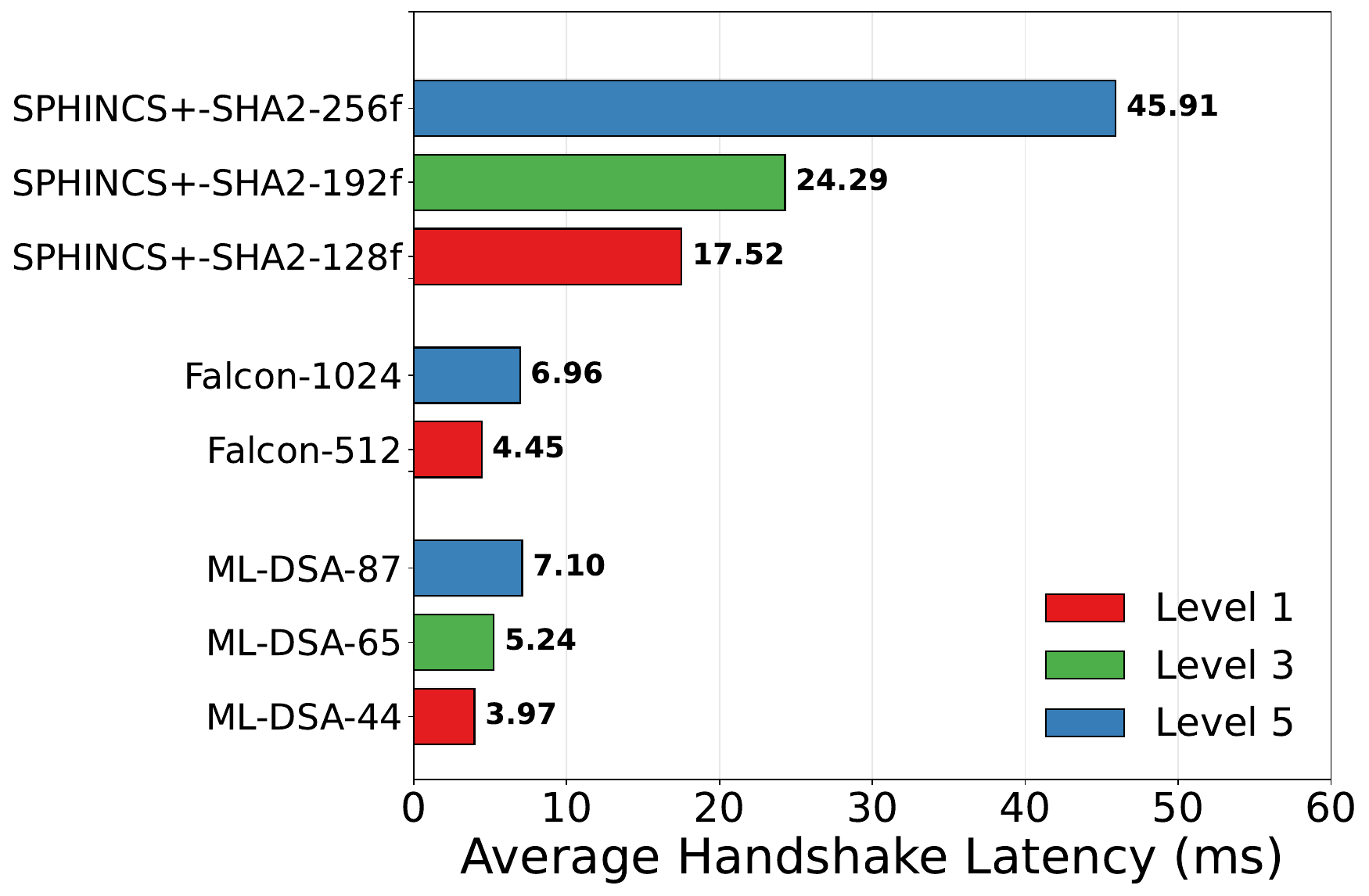}
    \caption{\footnotesize Handshake latency in data plane}
    \label{CPU:Sub1}
\end{subfigure}
\hfill
\begin{subfigure}[b]{0.445\textwidth}
    \centering
    \includegraphics[width=\textwidth]{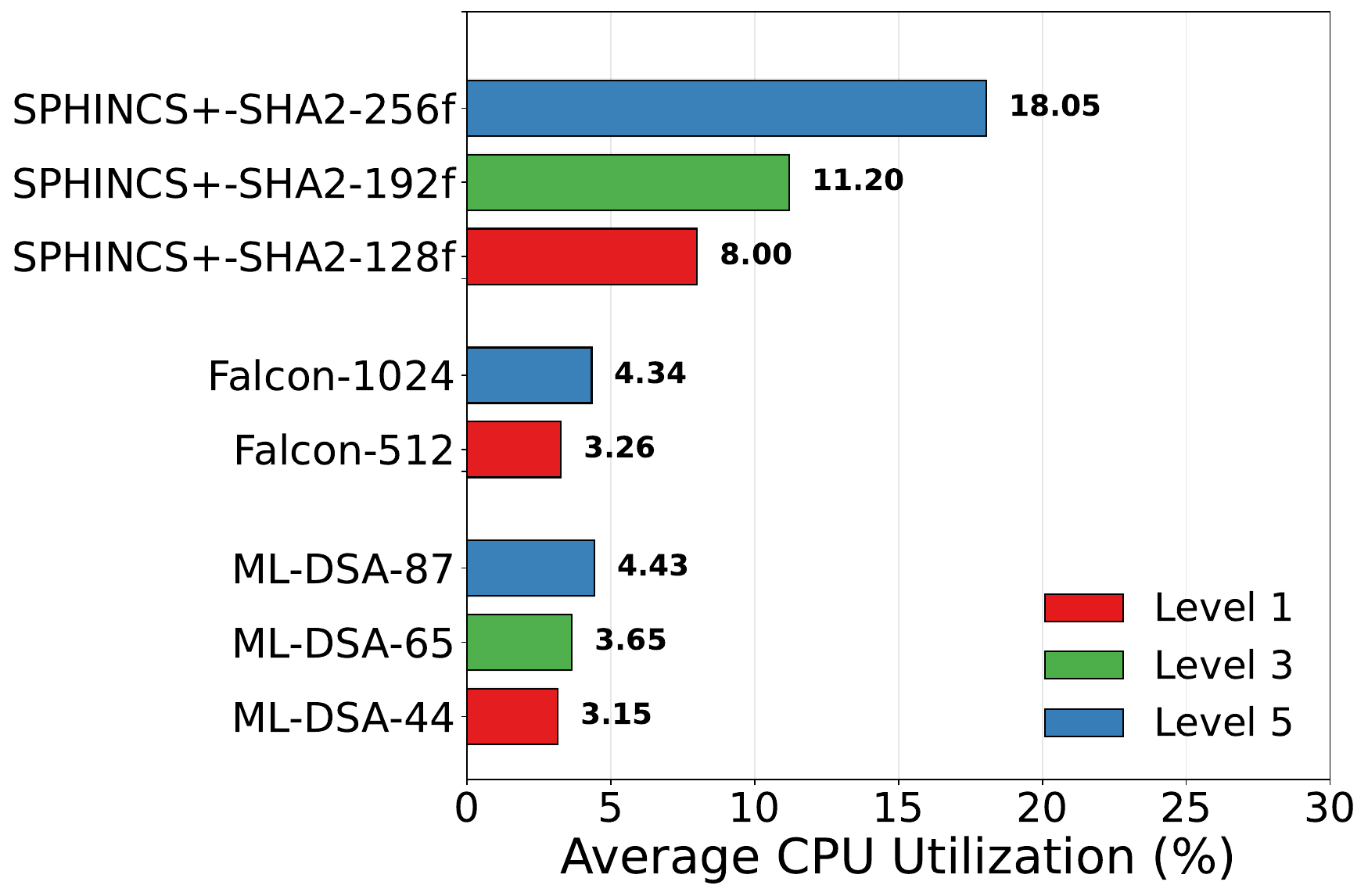}
    \caption{\footnotesize CPU utilization in control plane}
    \label{CPU:Sub2}
\end{subfigure}
\hfill
\caption{Comparing ML-DSA with SPHINCS+ and Falcon across (FrodoKEM is used for key establishment)}
\label{FrodoKEMFig}
\end{figure*}

\begin{figure*}[t!]
\centering
\begin{subfigure}[b]{0.445\textwidth}
    \centering
    \includegraphics[width=\textwidth]{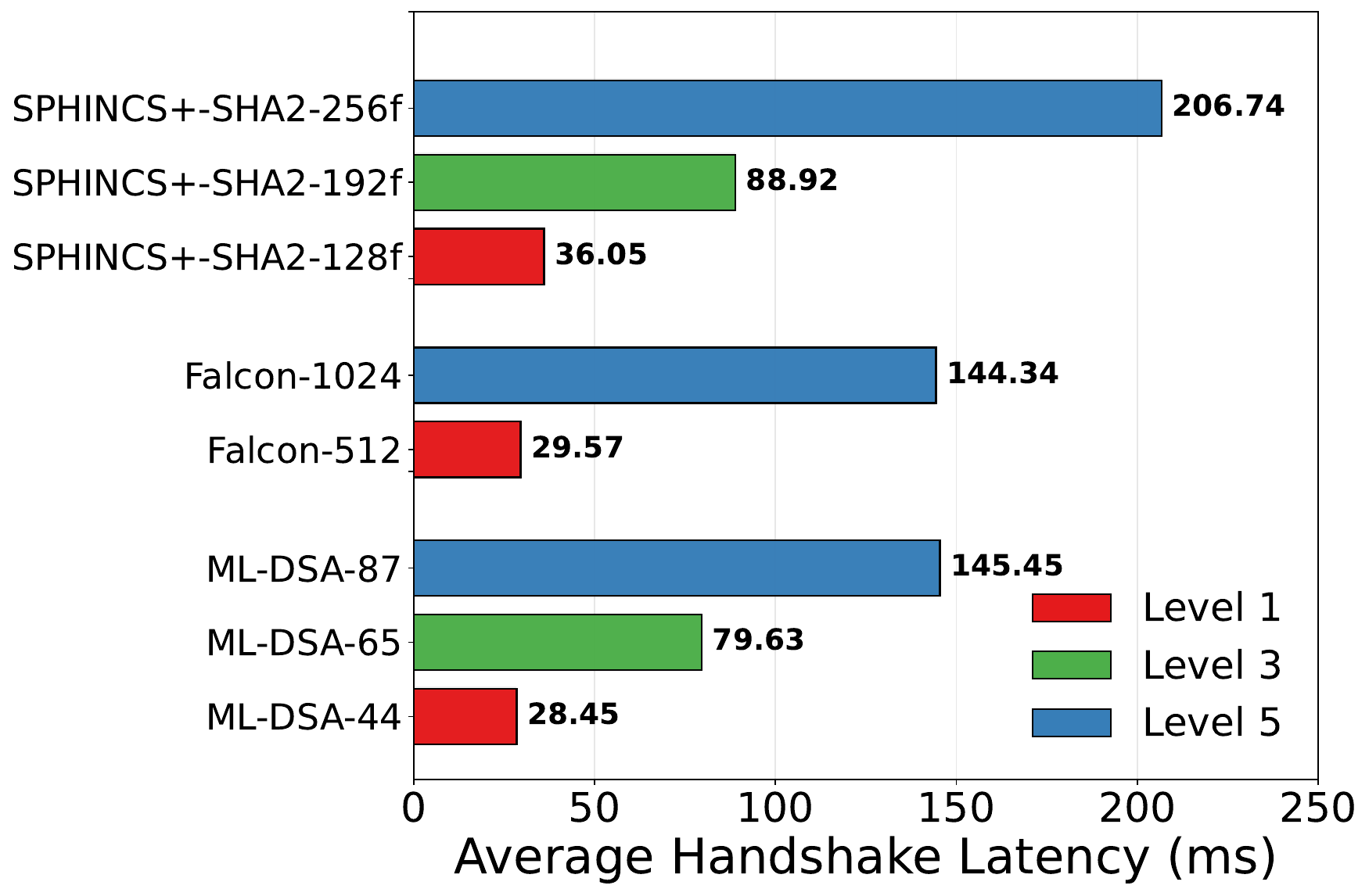}
    \caption{\footnotesize Handshake latency in data plane}
    \label{CPU:Sub1}
\end{subfigure}
\hfill
\begin{subfigure}[b]{0.445\textwidth}
    \centering
    \includegraphics[width=\textwidth]{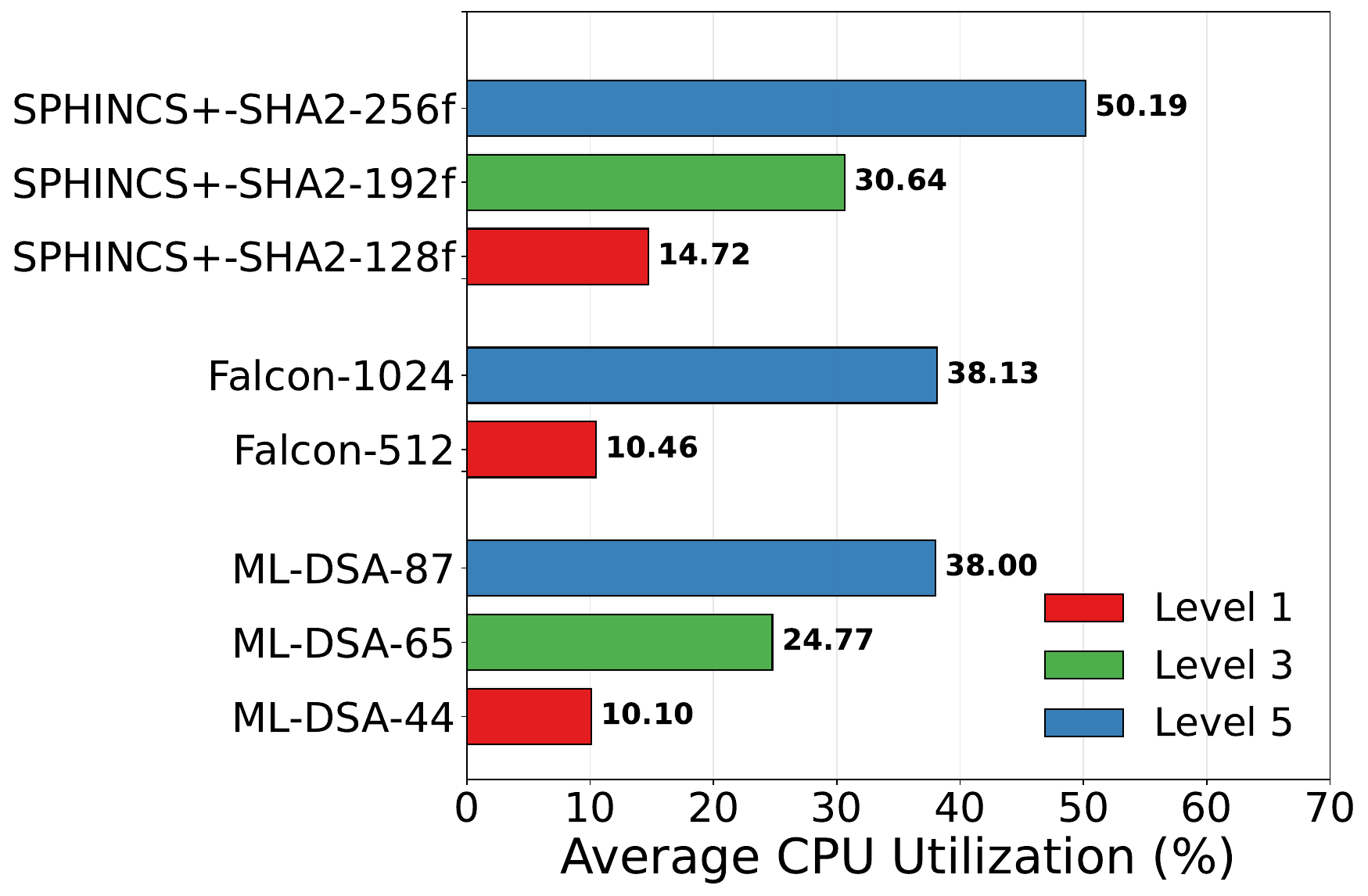}
    \caption{\footnotesize CPU utilization in control plane}
    \label{CPU:Sub2}
\end{subfigure}
\hfill
\caption{Comparing ML-DSA with SPHINCS+ and Falcon (HQC is used for key establishment)}
\label{HQCFig}
\end{figure*}

\textbf{Case 1 - ML-KEM}
As shown in Fig. \ref{SPHINCS}, ML-DSA and Falcon outperform SPHINCS+ in terms of latency and CPU utilization. In addition, Falcon-1024 shows the lower latency compared to ML-DSA-87. In addition, using ML-KEM for key establishment results in higher latency and CPU utilization compared to BIKE, FrodoKEM and HQC (See Fig. \ref{BikeFig}, \ref{FrodoKEMFig}, and \ref{HQCFig}).

\textbf{Case 2 - BIKE}
As shown in Fig. \ref{BikeFig}, using BIKE for key establishment results in higher latency and CPU utilization compared to ML-KEM. However it has slightly better results compared to FrodoKEM ( Fig. \ref{BikeFig}) and HQC (Fig. \ref{HQCFig}).

\textbf{Case 3 - FrodoKEM:}
As shown in Fig. \ref{FrodoKEMFig}, using FrodoKEM for key establishment results in higher latency and CPU utilization compared to ML-KEM and BIKE. However, it achieves better results in terms of latency and CPU utilization compared to HQC (Fig. \ref{HQCFig}).

\textbf{Case 4 - HQC}
As shown in Fig. \ref{HQCFig}, the worst results are achieved using HQC in terms of latency and CPU utilization. Overall, the results show that the choice of digital signature and key establishment methods affects the performance. For L1 and L3, the best results are achieved by combining ML-DSA and ML-KEM. For L5, Falcon achieves slightly better results compared to ML-DSA. The worst results are achieved with SPHINCS+.

\section{Conclusion}
In this work, we provide a proof-of-concept for a southbound API with PQ-TLS for secure SDN. Overall, the results show that for high security levels pure PQ-TLS, based on ML-DSA and ML-KEM, has better results in terms of latency and CPU utilization compared to legacy TLS and hybrid PQ-TLS. We compare the performance of ML-DSA with SPHINCS+ and Falcon. We also measure the performance with different key establishment schemes. For L1 and L3, the best results are achieved when ML-DSA and ML-KEM are used. For L5, the selection of Falcon and ML-KEM results in slightly better results compared to the selection of ML-DSA and ML-KEM.

\bibliographystyle{ACM-Reference-Format}
\bibliography{sample-base}

\end{document}